\documentclass[12pt]{article}

\pdfoutput=1

\usepackage[top=80pt,bottom=85pt,left=85pt,right=85pt]{geometry}
\usepackage{amssymb}
\usepackage{amsmath}
\usepackage{setspace}
\usepackage{sectsty}
\usepackage{cite}
\usepackage{bbm}
\usepackage[utf8]{inputenc}
\usepackage[T1]{fontenc}
\usepackage{comment}
\usepackage[english]{babel}
\usepackage{afterpage}
\usepackage{bbold}
\usepackage[usenames]{xcolor}
\usepackage{graphicx,subcaption}
\usepackage{setspace}
\usepackage{float}
\usepackage{booktabs}
\usepackage[debug,pageanchor=false]{hyperref}
\definecolor{link}{rgb}{.8,.15,.1}
\definecolor{pigment}{rgb}{0.36, 0.54, 0.66}
\definecolor{pigment2}{rgb}{0.19, 0.55, 0.91}
\definecolor{pigment3}{rgb}{0.2, 0.2, 0.6}
\definecolor{light-gray}{gray}{0.75}
\hypersetup{colorlinks=true,linkcolor=link,citecolor=link,urlcolor=link,linktocpage}

\usepackage{array,multirow,booktabs,longtable}
\usepackage{mathrsfs}
\usepackage{tikz-cd} 
\usetikzlibrary{backgrounds, arrows,calc,shapes,decorations.pathreplacing, decorations.markings, automata,positioning}
\tikzset{%
  >={Latex[width=2mm,length=2mm]},
            base/.style = {rectangle, rounded corners, draw=black,
                           minimum width=4cm, minimum heigwht=1cm,
                           text centered, font=\sffamily},
  activityStarts/.style = {base, fill=orange!15},
       startstop/.style = {base, fill=orange!15},
    activityRuns/.style = {base, fill=orange!15},
         process/.style = {base, minimum width=2.5cm, fill=orange!15,
                           font=\ttfamily},
}

\newcommand{\red}[1]{}

\tikzset{
        cvertex/.style={circle,draw=black,inner sep=1pt,outer sep=3pt},
        vertex/.style={circle,fill=black,inner sep=1pt,outer sep=3pt},
        star/.style={circle,fill=yellow,inner sep=0.75pt,outer sep=0.75pt},
        tvertex/.style={inner sep=1pt,font=\scriptsize},
        gap/.style={inner sep=0.5pt,fill=white}}

\tikzstyle{mybox} = [draw=black, fill=blue!10, very thick,
    rectangle, rounded corners, inner sep=10pt, inner ysep=20pt]
\tikzstyle{boxtitle} =[fill=blue!50, text=white,rectangle,rounded corners]

\newcommand{\cc}{\mathbb{C}}
\newcommand{\zz}{\mathbb{Z}}
\newcommand{\pp}{\mathbb{P}}

\def\cO{\mathcal{O}}

\DeclareMathOperator{\diag}{diag}

\newcommand{\todo}[1]{}
\renewcommand{\todo}[1]{{\color{red} TODO: {#1}}}
\renewcommand{\red}[1]{{\color{red} {#1}}}

\newcommand{\be}{\begin{equation}}  
\newcommand{\ee}{\end{equation}}  
\newcommand{\bea}{\begin{align}}
\newcommand{\eea}{\end{align}}
\newcommand{\bp}{\begin{bmatrix*}[r]}  
\newcommand{\ep}{\end{bmatrix*}}  
\newcommand{\bpp}{\begin{bmatrix}}  
\newcommand{\epp}{\end{bmatrix}}  
\newcommand{\bcd}{\begin{center}
\begin{tikzcd}}
\newcommand{\ecd}{\end{tikzcd} \end{center}}
\newcommand{\bm}{\begin{pmatrix}}  
\newcommand{\eem}{\end{pmatrix}}

\makeatletter
\@addtoreset{equation}{section}
\makeatother

\begin{document}


\begin{titlepage}

\begin{center}

\vskip .3in \noindent

{\Large \bf{Genus zero Gopakumar-Vafa invariants \\   from open strings}}

\bigskip\bigskip

Andr\'es Collinucci$^a$, Andrea Sangiovanni$^{b}$ and Roberto Valandro$^{b}$ \\

\bigskip


\bigskip
{\footnotesize
 \it

$^a$ Service de Physique Th\'eorique et Math\'ematique, Universit\'e Libre de Bruxelles and \\ International Solvay Institutes, Campus Plaine C.P.~231, B-1050 Bruxelles, Belgium\\
\vspace{.25cm}
$^b$ Dipartimento di Fisica, Universit\`a di Trieste, Strada Costiera 11, I-34151 Trieste, Italy \\
and INFN, Sezione di Trieste, Via Valerio 2, I-34127 Trieste, Italy	
}

\vskip .5cm
{\scriptsize \tt collinucci.phys at gmail.com \hspace{0.5cm}  andrea.sangiovanni at phd.units.it \hspace{0.5cm}  roberto.valandro at ts.infn.it }

\vskip 1cm
     	{\bf Abstract }
\vskip .1in
\end{center}
We propose a new way to compute the genus zero Gopakumar-Vafa invariants for two families of non-toric non-compact Calabi-Yau threefolds that admit simple flops: Reid's Pagodas, and Laufer's examples. We exploit the duality between M-theory on these threefolds, and IIA string theory with D6-branes and O6-planes. From this perspective, the GV invariants are detected as five-dimensional open string zero modes. 
We propose a definition for genus zero GV invariants for threefolds that do not admit small crepant resolutions. We find that in most cases, non-geometric T-brane data is required in order to fully specify the invariants.


\noindent

\vfill
\eject

\end{titlepage}


\tableofcontents

\newpage 
\section{Introduction} 
\label{sec:intro}
M-theory on non-compact singular Calabi-Yau threefolds is a well-known class of setups where five-dimensional SCFT's can be constructed.
That subject was initiated in the nineties in the two seminal papers \cite{Intriligator:1997pq,Seiberg:1996bd}. 
In the past five years, the subject has seen a revival, with more thorough systematic studies of such theories, their global symmetries, their moduli spaces, their prepotentials, and various methods for constructing them. It is difficult at this point to reference all works, some key works include \cite{Closset:2020afy,vanBeest:2020civ,Apruzzi:2019vpe,Bhardwaj:2018vuu,Collinucci:2008pf,Apruzzi:2019opn,Apruzzi:2019enx, Bhardwaj:2019jtr}.

The BPS spectrum of M-theory on CY threefolds is known to be captured by the A-model topological string \cite{Witten:1988xj}. More precisely, M2-branes wrapping holomorphic curves in the threefold are expected to give rise to particle-like states in the five-dimensional effective theory. Depending on the moduli space of one such curve, the kind of supermultiplet will vary. The simplest situation is a rigid curve, for which the effective multiplet is a hypermultiplet.

The Gopakumar-Vafa \cite{Gopakumar:1998ii, Gopakumar:1998ki, Gopakumar:1998jq} reformulation of the topological string makes this counting manifest, by recasting the partition function of the worldsheet as a generating function for integers known as GV invariants, or BPS invariants $n^g_{\beta}$, where $g$ is the genus of a curve, and $\beta \in H_2(X_3)$ its homology class.

For toric geometries, there are well-know systematic methods for computing these, such as the topological vertex \cite{Aganagic:2003db}. However, toric threefolds comprise but a subset of all possibilities. Even for rank-zero 5d SCFT's, most cases cannot be described torically.

A particularly interesting class of threefolds are those obtained as families of local K3's. One starts with a Du Val surface of ADE-type, parametrizes its versal deformations in an appropriate way, and takes a one-parameter family. This is known to give rise to a threefold with Gorenstein singularities, \cite{Katz:1992aa}.
For such types, techniques are known for finding their GV invariants, as pioneered in \cite{Katz:aa}, and further developed in more recent mathematical literature, \cite{Donovan:2013aa,Donovan:2015aa,Toda:2014aa}.

In this paper, we will study threefolds that are one-parameter families of A-type and D-type that admit simple flops. `Simple flops' means that these threefolds admit small resolutions such that only one exceptional $\pp^1$ is produced. 
We recast the problem of computing their GV invariants in terms of IIA string theory with D6-branes and O6-planes. This is a new way of looking at the curve counting problem in terms of open string calculations. 

For the A-series, we will tackle the so-called family of Reid's \emph{pagodas}. These are a class of conifold-like threefolds, which, despite being simple flops, can have arbitrarily high $n^{g=0}_{\beta}$ numbers. We will fiberwise reduce M-theory on such threefolds to IIA on $\cc^2$ with intersecting D6-branes, and in that setup, we will capture these invariants as an Ext$^1$ computation of open string spectra. We will also study a simple case, the \emph{generalized conifold}, which admits a double flop.

For the D-series, we will study, among others, the family of \emph{Laufer's examples} \cite{laufer}. These are the so-called `flops of length two'. The `length two' here means that, even though the threefold admits only a simple flop, the exceptional $\pp^1$ will allow for bound states of two M2-branes to wrap it. This will give rise to degree two GV invariants. We will recast this information in terms of IIA string theory with D6-branes and O6-planes. The orientifolding will allow for charge-two open string states, which will be shown to match the degree-two GV invariants of the geometry.

Finally, we will study CY threefolds that do not admit K\"ahler crepant resolutions. These are an interesting class of examples, because one cannot directly define their GV invariants, since no K\"ahler holomorphic curve can be produced. Nevertheless, they can be circle-reduced to IIA string theory with D6-branes that wrap singular non-compact divisors of $\cc^2$. From this perspective, we can compute the open string spectra. This leads us to define a notion of GV invariants for singular threefolds.
Furthermore, we find that these threefolds require additional non-geometric data in order to fully specify the low-energy spectrum, namely \emph{T-brane data}\cite{Cecotti:2010bp}. In IIA, these are vev's for open strings. In M-theory, they correspond to coherent states of M2-branes wrapped on the non-resolvable curve. This data is invisible to the CY geometry, but nevertheless impacts the GV invariants.
 

\section{GV invariants} \label{sec:bckndmat}

The topological A-model \cite{Witten:1988xj} is a simplified model of closed strings that counts holomorphic maps from the worldsheet into a Calabi-Yau threefold target space. Such maps are not always easy to count, as their images are sometimes holomorphic curves in the target space, but sometimes they correspond to points, or to multicovered curves. 

The total free energy for the A-model topological string is a genus-sum of the form
\be
F_{TF}({\bf t}) = \sum_{g=0} F_g({\bf t}) g_s^{2-2 g}\,,
\ee
where $t = B+i J$ is the complexified K\"ahler modulus, $F_g(t)$ is the genus-$g$ partition function, and $g_s$ is the string coupling constant. This sum splits 
\be
F_{TF}({\bf t}) = F_p({\bf t})+F_{WS}({\bf t})
\ee
into a perturbative part $F_{p}(t)$ and a worldsheet instanton part $F_{WS}(t)$. The perturbative part is a cubic polynomial in $t$, and the non-perturbative part is a sum over exponentials. Defining a basis 
\be
[c_i] \in H^{\rm cpt}_2(X_3, \zz)\,,
\ee
for the second homology with compact support, we have a sum over genera and general curve classes $\beta  = \sum_i d_i [c_i]$,
for a generic K\"ahler class ${\bf t} = \sum_i t_i [c_i]^{\vee}$\,,
where $\int_{c_i}[c_j^{\vee}] = \delta_i^j$. Defining a vector ${\bf d} = [d_1, \ldots, d_s]$, we can write the non-perturbative sum as follows:
\be
F_{WS}({\bf t}) = \sum_{g\geq 0} \sum_{{\bf d}}  N^{\bf d}_g e^{-{\bf d}\cdot {\bf t}}
\ee
where $N^d_g$ is the Gromov-Witten invariant at degree $d$ and genus $g$, which computes the virtual dimension of the moduli space of holomorphic maps from the worldsheet into the target space.

In the reinterpretation of Gopakumar and Vafa, \cite{Gopakumar:1998ii,Gopakumar:1998ki,Gopakumar:1998jq}, one looks not at the topological string, but at M-theory on the same CY threefold. The non-perturbative part of the free energy is recast into an object that counts curves in the target space. It takes the following form
\begin{align} \label{gvpartition}
F_{GV} ({\bf t}, g_s) &= \sum_ {w=1}^{\infty} \sum_{g=0}^{\infty} \sum_{{\bf d}}\frac{n^g_{\bf d}}{w} (2 \sin(w g_s/2))^{2g-2} e^{-w {\bf d}\cdot {\bf t}}\,.
\end{align}
This function now has a dependence on $g_s$. Formally
\be
F_{GV}({\bf t}, g_s) = F_{WS}({\bf t}, g_s)\,.
\ee
Here, we are summing over homology classes $\beta \in H_2^{\rm cpct}(X_3, \zz)$ with compact support, over \emph{degrees} $d \in \zz$, and genera $g$. Here, the degree represents a multiplicity of a homology class. The numbers $n^g_\beta$ are the Gopakumar-Vafa invariants, which are conjectured to be integers. 

The conjecture stems from the fact that, in M-theory, these integers are counting BPS states that are realized as M2-branes wrapping holomorphic curves. A higher degree means that there are M2-branes that wrap a given curve multiply.
As a separate phenomenon, notice that $\beta$ runs over all classes, and can therefore also run over multiples of a generator of $H_2$. Those are interpreted as bound states of coincident M2-branes wrapping a curve. Notice that we might have multi-wrapping, without there being bound states. In this paper, we will see bound states.

Each M2-state wrapping a holomorphic curve gives rise to a single particle state in the 5d effective field theory. In order to deduce what kind of super-multiplet describes such a particle, we refer to Witten's analysis, \cite{Witten:1996qb}, which extracts the spin of the lowest component of the superfield by studying the moduli space of the holomorphic curve. The upshot for the purposes of this paper is that a rigid curve gives rise to a 5d hypermultiplet. By `rigid', we mean either a curve with a negative normal bundle, or one with some higher order obstruction, such that its moduli space is a point. We will study situations with normal bundles $\mathcal{N} = \cO(-1) \oplus \cO(-1)\,; \cO(0) \oplus \cO(-2)\,;$ and $\cO(1) \oplus \cO(-3)$.
In all cases, the curves will be rigid, so we will only have hypermultiplet content in our 5d theories.

\section{Warmup case: The conifold}
The conifold is the simplest geometry that admits a `simple flop'. This means that it is a singular variety, and the `small resolution' grows a single $\pp^1$ as its exceptional locus. From the M-theory perspective, this sphere can support a wrapped M2-brane. Because the sphere has normal bundle $\cO(-1) \oplus \cO(-1)$, its moduli space is a point. From the analysis of \cite{Witten:1996qb}, we deduce that this corresponds to a hypermultiplet in the 5d theory. 

The resolved threefold admits one non-compact divisor that intersects the exceptional $\pp^1$ at one point. If the conifold is embedded into a compact CY threefold, then this divisor can be regarded as the Poincar\'e dual to a two-form, on which the supergravity $C_3$-form can be reduced, giving rise to a photon in 5d. In that case, the hypermultiplet would be charged under this photon, and we would have SQED with one flavor.
However, in the non-compact setting, such a two-form is non-normalizable (unless we specifically insist on having an ALF-type metric, see \cite{Collinucci:2020jqd}), in which case the $U(1)$ becomes ungauged, and is to be regarded as a flavor symmetry. 

The Gopakumar-Vafa partition function for the topological string on the conifold is known to be given by the following formula:
\begin{align}
F(t, g_s) &=\,\,\, \sum_{w=1}^{\infty}\frac{1}{w (2 \sin(w g_s/2))^2} e^{-w t}\,.
\end{align}
Comparing with \eqref{gvpartition}, we conclude that there is only one non-zero GV invariant at genus zero and degree one: $n^{g=0}_{d=1}=1$.
Here, $H_2(X, \zz) = \zz$ is generated by the class $[c]$ of the single exceptional $\pp^1$ and all its multiples $d[c]$.

In terms of the 5d effective field theory, we say that 
\be
n^0_{d} = \# \text{hypers with charge $d$ under the $U(1)$ flavor group.}
\ee
Hence, we deduce that there is one hyper with charge one. 

How do we recover this information from the type IIA string theory perspective? Let us define the conifold as the hypersurface:
\be
u v = z^2-w^2 \quad \subset \quad \cc^4\langle u, v, z, w \rangle\,.
\ee
This space admits several $\cc^* \cong S^1 \times \mathbb{R}$-actions. Let us choose the one that acts as follows:
\be
(u, v, z, w) \mapsto (\mu u, \mu^{-1} v, z, w)\,.
\ee
The $S^1$ part will be regarded as the M-theory circle. Reducing to IIA, we will have as a target space $\cc^2\langle z_1, z_2 \rangle \times \mathbb{R}$, where the real factor comes from the $\cc^*$-action. The D6-branes will be located at the loci where this action degenerates, i.e. at the reducible locus $(z+w)(z-w)=0$. So we have two flat D6-branes intersecting over a five-dimensional spacetime. A bifundamental string stretches between them, forming a free hypermultiplet.

We will now show a direct way to link the M-theory geometry to the IIA D6-brane configuration in terms of coherent sheaves. This viewpoint is a well-known story that has been expounded upon in many papers, but in \cite{Collinucci:2014qfa}, they are succinctly reviewed and tailor made for the applications we have in mind.

The central idea consists in describing the D6-branes in terms of \emph{tachyon condensates} of D8/anti-D8 pairs. Differently put, if we describe the D6-branes as coherent sheaves in $\cc^2$, then we can more readily extract the spectra. For this particular case, this is overkill, but in the rest of the paper, it will simplify calculation incredibly. 

We direct the reader to \cite{Collinucci:2014qfa} for a concise introduction to this language. Nevertheless, we will venture an even shorter review of the necessary concepts here.

Starting with our target space $X$ (in this case, $X := \cc^2$), we wish to describe D6-branes on a holomorphic subvariety $D \subset  X$, which may be reducible and non-reduced, equipped with a vector bundle $F_{DBI}$ over $D$. The way to describe this is to start with a pair of vector bundles $\tilde E$ and $E$ defined over all of $X$, and a linear bundle map 
\be
T: \tilde E \longrightarrow E
\ee
referred to as the \emph{tachyon map}. Here, $E$ is interpreted as a stack of D8-branes wrapping all of $X$, and $\tilde E$ is a stack of anti-D8-branes. $T$ is a matrix that encodes a set of bifundamental strings going from the anti-D8's to the D8's. 

If $T$ acquires a vev, then we say that tachyon condensation is taking place, and we get brane-anti-brane annihilation. If $T$ is the identity matrix, then we have total annihilation, and nothing remains. However, if $T$ is a holomorphic matrix, it will typically have complex codimension one loci where it fails to be invertible. At such points, annihilation does not take place, and we are left with D6-branes. These can be thought of as vortex solutions of a system over $X$. In mathematical terms, the cokernel of the map $T$ defines an object with support over the loci where $T$ is not invertible. This is encoded via a short exact sequence as follows:
\be
\begin{tikzcd}
0 \rar &  \tilde E \arrow{r}{T} & E \rar & \imath_{*} (F_{DBI}) \rar & 0\,,
\end{tikzcd}
\ee
where $\imath:D6 \hookrightarrow \cc^2$ is the embedding of the D6-worldvolume into the target space, and $\imath_{*} (F_{DBI})$ is the pushforward of the vector bundle over the brane into the target space. From now on, we will not display exact sequences, but simply the relevant part of the complexes. In this case, we will simply say that this D6-brane is given by the two term complex:
\be
\begin{tikzcd}
\tilde E \arrow{r}{T} & E\,.
\end{tikzcd}
\ee
Now we should consider the fact that there are gauge transformations on the D8 and the anti-D8 stacks, which in turn act bifundamentally on the tachyon field:

\begin{equation} \label{BifundTransfT}
\begin{tikzpicture}[scale=2]
\node (E) at (0,0) {$\tilde E$};
\node[right = of E] (F)   {$E$};
\node[right = of F] (ar) {$\Longrightarrow$};
\node[right = of ar] (T) {$T$};
\node[right = of T] (Tprime) {$G_{\rm D8}\, \cdot T \cdot \, G_{\overline{\rm D8}}^{-1}$};

 \draw (E) edge[->] node[above, font=\scriptsize]{$T$} (F);
 \draw (T) edge[->] (Tprime);
 \draw(E) edge[loop below] node[font=\scriptsize]{$G_{\overline{\rm D8}}$} (E) ;
 \draw(F) edge[loop below] node[font=\scriptsize]{$G_{\rm D8}$} (F) ;
\end{tikzpicture} 
\end{equation}
The gauge symmetry on the D6-brane system is given by the subset of transformations $(g_{\overline{\rm D8}}, g_{\rm D8})$ that leave the tachyon map $T$ invariant.

Having defined D6-branes and their gauge symmetries, we are now in a position to discuss open string spectra. Mathematically, it is defined as the first self-extension group of the coherent sheaf of the brane, Ext$^1\big(\imath_{*} (F_{DBI}), \imath_{*} (F_{DBI})\big)$. Concretely it is defined as the vertical map $\delta \phi$:
\be
\begin{tikzcd}
&\tilde E \dar{\delta \phi} \rar{T} & E\\
\tilde E \rar{T} & E
\end{tikzcd}
\ee
modulo gauge transformations. If we implement the transformations, it looks as follows:
\be
\begin{tikzcd}[row sep=huge, column sep=large]
&\tilde E \dlar[dashed]{g_{\overline{\rm D8}}} \dar{\delta \phi} \rar{T} & E \dlar[dashed]{g_{\rm D8}}\\
\tilde E \rar{T} & E
\end{tikzcd}
\ee
The `fluctuation' field $\delta \phi$ is  defined up to linearized gauge transformations as follows:
\be
\delta \phi \sim \delta \phi +T\cdot g_{\overline{\rm D8}} + g_{\rm D8} \cdot T\,.
\ee
These are referred to as \emph{homotopies} in mathematical language. 

Depending on the situation, we can create coincident, intersecting, or so-called \emph{T-branes} with this technology. All three will give rise to different kinds of spectrum. Let us proceed with our example step by step.

First, we build a pair of coincident D6-branes.\\ We start with two D8-anti-D8 pairs, and a \emph{tachyon map}, as a two-term complex, as follows:
\be
\begin{tikzcd}
\cO^{\oplus 2} \arrow{r}{T} & 
\cO^{\oplus 2}\,,
\end{tikzcd}
\ee
with the choice $T = z \cdot \mathbb{1}_2$, and $\cO$ is the \emph{structure sheaf} (trivial line bundle). The tachyon fails to be invertible at $z=0$. The gauge transformation pairs $(g_{\overline{\rm D8}}, g_{\rm D8})$ that leave this tachyon invariant must satisfy 
\be
g_{\overline{\rm D8}}=- g_{\rm D8}\,, \quad {\rm with} \quad g_{\rm D8} \in sl(2)_{\perp}\,.
\ee
Initially, we had a $sl(2) \oplus sl(2)$ gauge algebra, since there was a pair of D8-branes and a pair of anti-D8-branes. The gauge transformation we just identified is one subalgebra $sl(2)_{\perp}$. The orthogonal one, defined by the diagonal embedding of $sl(2)_{\Delta}$
\be
g_{\overline{\rm D8}}=+ g_{\rm D8}\,,
\ee
gives us our transformation law for the $\delta \phi$ fluctuations as follows:
\begin{align}
\delta \phi & \quad\sim \quad \delta \phi +T\cdot g_{\overline{\rm D8}} + g_{\rm D8} \cdot T = \delta \phi +z g\,,
\end{align}
with $g \cong \tfrac{1}{2} g_{\rm D8} \in sl(2)_{\Delta}$. Here, $\delta \phi \in sl(2)$, so this equivalence tells us that we can eliminate all dependence on $z$ in the matter field. In other words
\be
\delta \phi \in sl(2) \otimes \cc[z]/(z \cdot sl(2)) \cong sl(2)\,.
\ee
This means that this matter field is localized on the divisor $z=0$, and is therefore a seven-dimensional adjoint complex scalar on the worldvolume of the two D6-branes.

Now we consider switching on an angle between the two D6-branes. This means choosing
\be
T = \bm z-w & 0 \\ 0 & z+w \eem\,.
\ee

\be
\begin{tikzcd}[row sep=large, column sep=large]
 &\cO^{\oplus 2} \arrow{r}{T} \dar{\delta \phi} \dlar[dashed]{g_{\overline{\rm D8}}} & 
 \cO^{\oplus 2} \dlar[dashed]{g_{\rm D8}} \\
\cO^{\oplus 2} \arrow{r}{T}  & 
\cO^{\oplus 2} 
\end{tikzcd}
\ee
The choice of pair $(g_{\overline{\rm D8}}, g_{\rm D8})$ that preserves the tachyon is
\be
 g_{\rm D8} = \bm g & 0 \\0 & -g \eem\,, \quad g_{\overline{\rm D8}}=- g_{\rm D8}
  \qquad g \in u_{\cc}(1)\,.\ee
This identifies a $u_{\cc}(1) \subset sl(2)_{\perp} \subset sl(2) \oplus sl(2)$, where $sl(2)_{\perp}$ is the `relative' linear combination defined earlier. 

The broken generators for the $(g_{\overline{\rm D8}}, g_{\rm D8})$ pair are then:
\be
g_{\overline{\rm D8}} = \bm \tfrac{1}{2} g & a_+\\ a_- & -\tfrac{1}{2} g \eem\,, \quad g_{\rm D8} = \bm \tfrac{1}{2} g & b_+\\ b_- & -\tfrac{1}{2} g \eem\,.
\ee
These induce the following gauge equivalence for the matter field $\delta \phi\equiv\bm \delta \phi_0 & \delta \phi_+\\ \delta \phi_- & -\delta \phi_0\eem$
\be
 \delta \phi \sim \delta \phi + \bm g (z-w) & (a_++b_+) z + (a_+-b_+) w \\ (a_-+b_-) z + (b_--a_-) w  & -g (z+w) \eem\,.
\ee
We defined the fluctuation as traceless, since we are not interested in center of mass motion. From this we learn that the $\delta \phi_0$ mode is localized in codimension one on the target space. In other words, it's a 7d field. On the other hand, the off-diagonal modes are localized in codimension two, and are hence legitimately dynamical from a 5d viewpoint:
\be
\delta \phi_\pm \in \cc[z, w]/(z, w) \cong \cc\,.
\ee
The IIA interpretation is that we have two intersecting flat D6-branes, and there is one 5d hypermultiplet $(\delta \phi_+, \delta \phi_-)$, depicted by the following quiver:
\be
\begin{tikzpicture}
\node (d61) {D6$_1$};
\node[right = 2cm of d61](d62){D6$_2$};
\draw (d61) edge[->, bend right] node[below]{$\delta \phi_-$} (d62) ;
\draw (d62) edge[->, bend right] node[above]{$\delta \phi_+$} (d61);
\end{tikzpicture}
\ee
Hence, there is a multiplicity one hypermultiplet of charge one w.r.t. the unbroken $U(1)$ flavor\footnote{This is a flavor symmetry for the 5d theory; it comes from a 7d gauge symmetry on the worldvolume of the D6-branes.}  symmetry.
In M-theory, this translates to the fact that, on the conifold, there is only the genus zero, degree one GV invariant, and it is equal to one:
\be
n^{g=0}_{d=1} = 1\,.
\ee

\section{Flops of length one}
\subsection{Reid's pagodas}
Now we will study Reid's `pagodas' \cite{pagodas}, which comprise a class of threefolds that admit simple flops, whereby the normal bundle for the exceptional $\pp^1$ is $\cO \oplus \cO(-2)$. The curves are obstructed, even though this is not apparent from the normal bundle. Hence, they still have zero-dimensional moduli spaces, thereby giving rise to hypers in the effective theory, following Witten's analysis.

We will define the `k-th Reid pagoda', RP$_k$ as follows:
\be
u v = (z^k-w) (z^k+w) \quad \subset \quad \cc^4\,.
\ee
This variety is not toric, and, just as the conifold can be though of as a family of $A_1$ surfaces, this one can be thought of as a fibration of deformed $A_{2 k-1}$ surfaces fibered over the $w$-plane. It can be small-resolved very similarly to the conifold, by introducing two projective coordinates $[s:t] \in \pp^1$ and imposing the equation
\be
\bm u & z^k-w \\z^k+w & v \eem \bm s\\t \eem = 0\,, \quad \subset \quad \cc^4 \times \pp^1\,.
\ee
Na\"ively, because there is only one exceptional $\pp^1$ for M2-branes to wrap, one might expect there to be a single hyper in the effective theory. However, we will see that this is not the case. 

Let us describe this setup from a IIA perspective as the following complex:
\be
\begin{tikzcd}
\cO^{\oplus 2} \arrow{r}{T} & 
\cO^{\oplus 2}\,,
\end{tikzcd}
\ee
with 
\be
T:= \bm z^k-w & 0\\0&z^k+w \eem\,.
\ee
This corresponds to two intersecting D6-branes on the divisors $z^k \pm w=0$, respectively. Note, however, that these two branes do not intersect transversely, but intersect at a multiplicity $k$ point, i.e. a \emph{non-reduced scheme}. In terms of ideals, we write that
\be
(z^k-w, z^k+w) = (z^k, w)
\ee
which is a `fat point' in $\cc^2$. Hence, we have no right to expect the spectrum to be one-dimensional. A first hint that this might be so is the fact that the slightest deformation of either brane will split up the fat point into $k$ distinct points:
\be
(z^k-w+\delta, z^k+w) = (z^k+\delta, w)\,.
\ee
Hence, we should anticipate $k$ hypermultiplets of charge one under the flavor $U(1)$. Let us compute this with the techniques explained in the previous section:

We have again five broken $sl(2)$ generators, meaning that they do not commute with the tachyon:
\be
g_{\overline{\rm D8}} = \bm \tfrac{1}{2} g & a_+\\ a_- & -\tfrac{1}{2} g \eem\,, \quad g_{\rm D8} = \bm \tfrac{1}{2} g & b_+\\ b_- & -\tfrac{1}{2} g \eem\,.
\ee
Now we compute how these generators act as linearized gauged transformations on a fluctuation field $\delta \phi\equiv\bm \delta \phi_0 & \delta \phi_+\\ \delta \phi_- & -\delta \phi_0\eem$:
\be
\delta \phi
\sim \delta \phi + \bm g (z^k-w) & (a_++b_+) z^k + (a_+-b_+) w \\ (a_-+b_-) z^k + (b_--a_-) w  & -g (z^k+w) \eem\,.
\ee
The $\delta \phi_0$ mode localizes to $w\pm z^k=0$, and is therefore a 7d field. From the 5d perspective, it is non-dynamical. The bifundamental modes, on the other hand, localize on the ideal
\be
(z^k, w)\,.
\ee
So we have that
\be
\delta \phi_{\pm} \in \cc[z, w]/(z^k, w) \cong \cc[z]/(z^k)\,.
\ee
This ring can be regarded as a $k$-dimensional vector space
\be
\cc[z]/(z^k) \cong \cc^k\,.
\ee
Therefore, we conclude that there are $k$ hypermultiplets of charge one. This matches the known result that Reid's pagodas have the following GV invariants:
\be
n^{g=0}_{d=1} = k\,,
\ee
and zero for all other classes and genera.

\subsection{Non-simple flops: The generalized conifold}
Let us now consider another interesting geometry that admits a non-simple flop of length one. That is to say, the small resolution gives rise to a collection of $\pp^1$'s, each with multiplicity one. It is dubbed `generalized conifold' in \cite{Aspinwall:2010mw}. The definition is the following hypersurface:
\be
uv = z^3-w^2 z \quad \subset \quad \cc^4\,.
\ee
This geometry admits the following small resolutions:
\begin{align}
\bm u & z \\ z^2-w^2 & v \eem \cdot \bm s_1\\t_1 \eem&=0 \qquad \subset \quad \cc^4 \times \pp^1_{[s_1: t_1]}\,,\\
\bm u & z (z-w) \\ z+w & v \eem \cdot \bm s_2\\t_2 \eem&=0  \qquad \subset \quad \cc^4 \times \pp^1_{[s_2: t_2]}\,.\\
\end{align}
Let's call the classes of these two $\pp^1$'s $\beta_1$ and $\beta_2 \in H_2(X_3, \zz)$.
We recognize a $\cc^*$-fibration from the $(u,v)$ pair, and deduce that there is a IIA system of D6-branes on the reducible divisor $z (z-w) (z+w)$. From this, we can venture the following tachyon profile:
\be
T = \bm z &0 & 0 \\ 0 & z-w &0 \\ 0 & 0 & z+w \eem\,.
\ee
The unbroken flavor symmetry (i.e. 7d gauge symmetry) here is $u_{\cc}(1)_1 \oplus u_{\cc}(1)_2$, generated by the Cartan of the subalgebra $sl(3)_{\perp} \subset sl(3) \oplus sl(3)$:
\be
-g_{\overline{\rm D8}}= g_{\rm D8} = \bm g_1 & 0 & 0 \\0 & g_2 & 0 \\ 0 & 0 & g_3 \eem\,; \quad {\rm with} \quad \sum_i g_i=0 \qquad \subset \quad u_{\cc}(1)_1 \oplus u_{\cc}(1)_2\,.
\ee
In terms of the open string spectrum, we see that the diagonal elements are localized only in 7d along $z=0$. On the other hand, the offdiagonals are easily seen to be localized at $(z, w)$. This can be derived by parametrizing the most general left and right transformations, as we did in the previous cases. More intuitively, though, we just need to realize that the left and right transformations can be used to implement Gaussian elimination moves on the tachyon matrix. 

For instance, we see that by adding a multiplet of the first column to the second column, and a multiple of the second row to the first row, we can gauge away any $z$ and $w$ dependence from the $(1, 2)$ entry. Proceeding in this way, we deduce that there are three hypermultiplets: $(H_{(1,-1,0)}, H_{(0,1,-1)}, H_{(1,0,-1)})$, where the subscripts denote the charges w.r.t. the three linearly \emph{dependent} $u_{\cc}(1)$'s. Note that the third hyper has the same charges as the product of the first two hypers. This is the IIA counterpart of the statement in M-theory that there are three possible M2-brane states: One wrapping the class $[c_1]$, one wrapping $[c_2]$, and one wrapping $[c_1]+[c_2]$.

From this, we predict the following genus zero GV invariants $n^g_{(d_1, d_2)}$:
\be
n^0_{(1,0)} = 1\,, \qquad n^0_{(0,1)} = 1\,, \qquad n^0_{(1,1)} = 1\,,
\ee
and all other invariants equal to zero.

We can generalize this analysis to the following class of geometries:
\be
u v = z^k-w^k\,.
\ee
First, we recognize that the r.h.s. can be factorized as follows:
\be
u v = \prod_{i=0}^{k-1} (z-\zeta^i w)\,, 
\ee
where $\zeta$ is the $k$-th root of unity ($\zeta^k=1$). Then, we define the following tachyon matrix:
\be
T = \diag(z-\zeta^0 w\,; z-\zeta^1 w\,; \ldots\,; z-\zeta^{k-1})\,.
\ee
From this, we deduce that the open string spectrum consists in filling the off-diagonal entries with functions that are localized at $(z, w)$, meaning they are five-dimensional hypers. The preserved flavor symmetry here is 
\be
u_{\cc}(1)^{\oplus k-1}
\ee
the Cartan subalgebra of an $sl(k)$ algebra. The hypers organize themselves into the roots of $sl(k)$, and, in the M-theory uplift, each root $\alpha$ translates to a class $[\alpha] \in H_2(X_3, \zz)$. Defining a basis $\langle \alpha_i \rangle$, of simple roots, we draw the conclusion that that
\be
n^0_{(n_1, \ldots, n_{k-1})} = 1\,, \quad \iff \quad {\sum_i n_i \alpha_i}\quad\text{is a root}\,.
\ee

\section{Flops of length two}
\subsection{Generalities}
In this section, we get to a class of threefolds that bare a starker contrast to the conifold than does the family of Pagoda's. There are threefolds that admit simple flops, insofar as the exceptional locus is still a single $\pp^1$. However, in contrast to the flops of length one, the \emph{scheme-theoretic multiplicity} is two. They are discussed at length in \cite{Curto:aa} and \cite{Aspinwall:2010mw}. 

In general terms, say we have a threefold admitting a simple flop of length $l$. Let $\tilde X_3$ be the resolved space, $X_3$ the singular threefold, and $p$ the singular point. In terms of the blowdown map $\pi: \tilde X_3 \longrightarrow X_3$, if we pullback the skyscaper sheaf of the singular point, the result contains a `fat' $\pp^1$. More explicitly:
\be
\pi: \tilde X_3 \longrightarrow X_3 \qquad \pi^*(\cO_p) = \cO_{\pp^1}^{\oplus l} \oplus \cal{S}\,,
\ee
where $\cal{S}$ is some object of the derived category of coherent sheaves on $\tilde X_3$. For the conifold, we would have simply had $\cO^{\oplus 1} \oplus \cO(-1)[1]$.

In terms of physics, this can be understood in the framework of branes probing the singularity. A point-like D-brane at $p$ will split into fractional branes, as understood in \cite{Douglas:1996sw}. The effective field theory will look like a quiver of the following form:
\be
\begin{tikzpicture}
\node (R) {$U(1)$} ;
\node[right=of R]  (L) {$U(l)$};
\draw (R) edge[->, bend right] (L);
\draw (L) edge[->, bend right] (R);
\draw[->] (L) edge[loop right] (L);
\draw[->] (R) edge[loop left] (R);
\end{tikzpicture}
\ee
whereby all arrows can have multiplicities.

These threefolds are constructed as families of $D_n$-type local K3 surfaces, whereby only a node of Dynkin weight two is kept small. What is interesting is the fact that this weight two leads us to expect there to be GV invariants of degree one and two w.r.t. the generating class $\{[\beta]\} \in H_2(\tilde X_3, \zz)$.
We refer to \cite{Curto:aa} and \cite{Karmazyn:2017aa} for the explicit construction methods. In the works \cite{Collinucci:2018aho,Collinucci:2019fnh}, these geometries and their small resolutions are described explicitly as quiver representations.

\subsection{Orientifolds}
In order to connect such threefolds to IIA constructions, we will need not only D6-branes, but also O6$^-$-planes. Intuitively, this makes sense, since O6$^-$-planes allow us to generate $D_n$ groups. Let us setup the general machinery, and proceed with examples in subsequent sections.

We take the target space where the D6-brane lives to be $\cc^2 \langle \xi, w \rangle$ and we define a $\zz_2$ orientifold action $\sigma$ on the target space as follows:
\be
\sigma: (\xi,  w) \longrightarrow (-\xi,  w)\,.
\ee
This leaves the divisor $\xi=0$ invariant. On the worldsheet, the action is $\Omega_p(-1)^{F_L}$, where $\Omega_p$ is the worldsheet parity and $F_L$ is the  space-time fermion number in the left-moving sector. 

Now we need to define the action of the orientifold projection on the tachyon. We refer to \cite{Collinucci:2008pf} for general details. The upshot is that a tachyon complex
\be
E \stackrel{T}\longrightarrow F
\ee
must obey the following:
\be
E \cong F^{\vee} \qquad {\rm and} \quad T = -\sigma^*(T^t)\,. \label{tachyoncond}
\ee
Consistency requires that left and right gauge transformations be related as follows:
\be	
T \longrightarrow G \cdot T \cdot \sigma^*(G^t)\,.
\ee

Our CY threefolds will be defined as $\zz_2$ quotients of $\cc^*$-fibrations over the double-cover IIA geometry. This is summarized in the following diagram:
\be\label{doubleCoverOr}
\begin{tikzcd}	
  \overline{X_3} \arrow[r, red, "\zz_2" blue] \arrow[d, "\cc^*" left]
    & X_3 \arrow[d, "\cc^*"] \\
  \overline{X_2} \arrow[r, red, "\zz_2" blue]
&\cc^2 \end{tikzcd}
\ee
Here, $\overline{X_2}$ is the double cover of $\cc^2$. This is the `upstairs' space where the orientifold action is reflecting the coordinate $\xi$. Its quotient is achieved as follows:
\begin{align}
\pi_{\zz_2}:& \overline{X_2}\longrightarrow \cc^2\\
& (\xi, w) \mapsto (z:=\xi^2, w)\,.
\end{align}
This means that we can describe $X_2$ redundantly as a hypersurface:
\be
X_2:= \quad \cc[\xi, w, z]/(\xi^2-z)\,.
\ee
From this perspective, all polynomials on $X_2$ can be split into an orientifold even and an odd part:
$\gamma(\xi, w, z) = \alpha(w, z)+ \xi \beta(w, z)$.
Any higher powers in $\xi$ can be eliminated with the hypersurface equation. Combining this with our condition on the tachyon \eqref{tachyoncond}, we arrive at the following general Ansatz:
\be\label{tachyon symmetry}
T = A+\xi S\,, \qquad {\rm with} \quad A^t = -A\qquad {\rm and} \quad S^t = S\,,
\ee
where the entries of $A$ and $S$ are polynomials in $\cc[w,z]$.

Let us now go to M-theory.
The action of $\Omega_p(-1)^{F_L}$ is realized in the M-theory uplift by the inversion on the extra circle. The circle belongs to the $\mathbb{C}^\ast$ fiber of threefold $\overline{X}_3$. If we represent the fiber on a generic point by $u v = c $, the circle inversion is realized by $u\leftrightarrow v$. Hence, we expect that the $\mathbb{C}^\ast$ fibration $\overline{X}_3$ is described by an equation like
\be
(x+\xi y)(x-\xi y) = P(w,\xi^2)  \qquad \subset \cc^4 \langle \xi, x, y, w \rangle \:.
\ee
Following the diagram \eqref{doubleCoverOr}, we expect for $X_3$ a $\cc^\ast$-fibration like
\be
x^2-zy^2 = P(w,z)  \qquad \subset \cc^4 \langle x, y, w, z \rangle \:.
\ee
If we are given such a form, we immediately read the location of the orientifold plane at $z=0$. Moreover the D6-brane locus is where the $\cc^\ast$ factorizes, i.e. where the quadric $zy^2-P$ (in $y$) becomes a square. This happens when the discriminant $\Delta=zP$ vanishes.\footnote{The most generic form of such a fibration is actually $x^2=zy^2+2Q(w,z)y - P(w,z)$ and the D6-branes sit on the discriminant locus $\Delta\equiv Q^2+zP=0$.}

\subsection{Laufer's examples}\label{Sec:Laufer}
The first examples of flops of length 2 were given by Laufer \cite{laufer}.
This particular class of length two flops is defined by the following hypersurfaces:
\begin{equation}
 x^2 - z y^2 - w \left ( w^2 - z^{2n+1} \right)  = 0 \qquad\mbox{with }n\geq 1 \:.
\end{equation}

This is a $\mathbb{C}^\ast$ fibration over the space $\mathbb{C}^2 \langle z,w \rangle$.
The $\mathbb{C}^\ast$ fiber degenerates into two $\mathbb{C}$ over the locus 
\begin{equation}
 \Delta \equiv w \,z \, \left(   z^{2n+1} - w^2 \right) = 0 \:.
\end{equation}
This is the D6-brane locus. Moreover the $\mathbb{C}^\ast$ fiber degenerates further to two coincident $\mathbb{C}$ on top of $z=0$, that is identified with the location of the O-plane.

The double cover  is given by $\xi^2 =  z$, where the orientifold involution is $\xi\mapsto - \xi$. The fixed point locus is correctly at $z=0$.
In the double cover twofold, the D6-brane locus is given by 
\begin{equation}\label{D6brlocusLaufer}
w\,\xi ^2 \left(\xi ^{2 n+1}+w\right)\left(\xi ^{2 n+1}-w\right) = 0
\end{equation}
and is given by the complex
\be
\begin{tikzcd}
\cO^{\oplus 4} \arrow{r}{T} & 
\cO^{\oplus 4}\,,
\end{tikzcd}
\ee
where $T$ is 
\begin{equation}\label{TachLauf}
T =\left(
\begin{array}{cccc}
 0 & \xi ^{2 n+1}+w & 0 & 0 \\
 \xi ^{2 n+1}-w & 0 & 0 & 0 \\
 0 & 0 & \xi  & 0 \\
 0 & 0 & 0 & w \xi  \\
\end{array}
\right)\:.
\end{equation}
Its determinant correctly reproduces the brane locus \eqref{D6brlocusLaufer}. 
Notice that the form of the matrix \eqref{TachLauf} is compatible with the orientifold invariance condition \eqref{tachyon symmetry}.
One may object that \eqref{D6brlocusLaufer} suggests that there are five D6-branes (one invariant brane at $w=0$, two branes on top of the O6-plane and a pair of one brane and its orientifold image), and then one would expect a $5\times 5 $ tachyon matrix. However there is no way to build a $5\times 5 $ matrix that respects \eqref{tachyon symmetry} and that reproduces \eqref{D6brlocusLaufer}. One entry of $T$ must be $\xi w$, which is interpreted as a bound state of two branes (see \cite{Cecotti:2010bp, Collinucci:2014qfa}).

The linearized $D8/\overline{D8}$ gauge transformations acts on $T$ as 
\begin{equation}\label{TtransformLaufer}
  T \mapsto g\cdot T + T \cdot \sigma^\ast g^t \:.
\end{equation}
where the orientifold invariance of the setup forces the relation $G_{\rm \overline{D8}}=\sigma^\ast G_{\rm D8}^t$.

The surviving gauge symmetry is again given by the $D8/\overline{D8}$ gauge transformations that leave $T$ invariant. In the present case we still have $U(1)$ gauge symmetry, whose generator is
\begin{equation}\label{U1generatorLaufer}
g_{U(1)} =\left(
\begin{array}{cccc}
 1 & 0 & 0 & 0 \\
 0 & -1 & 0 & 0 \\
 0 & 0 & 0  & 0 \\
 0 & 0 & 0 & 0  \\
\end{array}
\right)
\end{equation}

We are now ready to compute the spectrum of zero modes. The elements of the $gl(4)$ $D8/\overline{D8}$ gauge symmetry that act non-trivially on $T$ are given by
\begin{equation}\label{BrokenGenLaufer}
g_{D8} =\frac12 \left(
\begin{array}{cccc}
 \alpha_{11}+\xi \beta_{11} & \alpha_{12}+\xi \beta_{12} & \gamma_{13} & \gamma_{14} \\
 \alpha_{21}+\xi \beta_{21} &  \alpha_{11}-\xi \beta_{11} &  \gamma_{23}  &  \gamma_{24}  \\
  \gamma_{31}  &  \gamma_{32} & \alpha_{33}  &   \alpha_{34}+\xi \beta_{34} \\
   \gamma_{41} &  \gamma_{42} &  \alpha_{43}+\xi \beta_{43} & \alpha_{44}  \\
\end{array}
\right)
\end{equation}
where $\alpha_{ij}$ and $\beta_{ij}$ are polynomials invariant under $\xi\mapsto - \xi$, while $\gamma_{ij}$ are generic polynomials in $w,\xi$.

The fluctuation field $\delta\phi$ (that is a matrix with entries $\delta\phi_{ij}$) is found by modding out the deformations that can be obtained by a linearized gauge transformation \eqref{TtransformLaufer}:
\begin{equation}
\delta \phi \sim \delta \phi + g_{D8} \cdot T +T\cdot \sigma^\ast g_{D8}^t \equiv \delta \phi + \delta_g T
\end{equation}
where we remind that $\sigma^{\ast}$ will force $\xi\mapsto -\xi$.

Let us write $\delta_gT$ in a block-diagonal form (with $2\times 2$ blocks):
\begin{equation}
\delta_g T = \left(
\begin{array}{c|c}
A & B \\
\hline
C & D \\
\end{array}
\right),
\end{equation}

Due to the block-diagonal form of the matrix $T$, the blocks $A,D$ of $\delta_g T$ are affected only by the corresponding $2\times2$ diagonal blocks of $g_{D8}$. Moreover the off-diagonal blocks of $\delta_gT$ are related by $C=-\sigma^\ast B^t$ and depend only on the parameters appearing in the off-diagonal blocks of $g_{D8}$. Hence, we can study separately the blocks of $\delta_g T$.

We now compute separately each block of $\delta_g T$ and we discuss which modes it can fix. We start with the diagonal block $A$: 
\begin{equation}
\setlength{\jot}{16pt} 
\begin{split}
&A= \left(
\begin{array}{cc}
 2\alpha_{12} \xi ^{2 n+1}-2 \beta_{12} w\,\xi & 2\left( \alpha_{11} + \beta_{11}\xi \right) \left(\xi^{2n+1}+w \right)  \\
 2\left( \alpha_{11} - \beta_{11}\xi \right) \left( \xi^{2n+1} - w \right)  & 2\alpha_{21} \xi ^{2 n+1}+2 \beta_{21} w\,\xi \\
\end{array}
\right)\:.\\
\end{split} 
\end{equation}
The modes $\delta\phi_{12}$ and $\delta\phi_{21}$ localize respectively to the D6-brane loci $w+\xi^{2n+1}$ and $w-\xi^{2n+1}$. They are 7d fields that are non-dynamical in 5d. Due to the orientifold projection condition \eqref{tachyon symmetry}, the bifundamental modes are of the form 
$$\delta\phi_{11}=\xi \delta\phi_{A+} \qquad \mbox{ and } \qquad \delta\phi_{22}=\xi \delta\phi_{A-}$$ 
with $\delta\phi_{A\pm}$  polynomials that are invariant under $\xi\mapsto -\xi$.
The modes $\delta\phi_{11}$ and $\delta\phi_{22}$ localize on the ideal
\begin{equation}
(w\xi\,,\,\, \xi^{2n+1})
\end{equation}
and then
\begin{equation}
  \delta\phi_{A\pm} \in \mathbb{C}[\xi,w]^{\rm inv}/(w,\, \xi^{2n}) \cong\mathbb{C}[z,w]/(w,\, z^{n}) \cong \mathbb{C}[z]/( z^{n}) \cong \mathbb{C}^n \:.
\end{equation}
We then conclude that there are $n$ hypermultiplets related to the up-left diagonal block of $\delta\phi$. They have charge $2$ under the $U(1)$ generated by \eqref{U1generatorLaufer}, as it can be shown by taking $g_{U(1)} \cdot \delta\phi + \delta\phi \cdot g_{U(1)} $. This can be understood by noticing that these modes come from strings stretching from one $U(1)$ brane to its image: they live in the symmetric representation of the group on the brane, hence they have charge 2 under the $U(1)$ group.

Let us move to the diagonal block $D$:
\begin{equation}
\setlength{\jot}{16pt} 
\begin{split}
& D = \left(
\begin{array}{cc}
 2 \alpha_{33}\xi & 
 \left( \alpha_{34}+\beta_{34}\xi \right)w\xi  + \left( \alpha_{43}-\beta_{43}\xi \right)\xi  \\
 \left( \alpha_{34}-\beta_{34}\xi \right)w\xi  + \left( \alpha_{43}+\beta_{43}\xi \right)\xi & 
 2 \alpha_{44}\xi w\\
\end{array}
\right) \\
\end{split} 
\end{equation}
Here we see that none of the modes $\delta\phi_{33},\delta\phi_{34},\delta\phi_{43},\delta\phi_{44}$ are localized in 5d.

Finally we analyse the block $B$ (the modes in the block $C$ are the orientifold image of the ones in $B$):
\begin{equation}
\setlength{\jot}{16pt} 
\begin{split}
& B = \left(
\begin{array}{cc}
\gamma_{13} \xi   + \sigma^\ast\gamma_{32} \left( w + \xi^{2n+1} \right)  & \gamma_{14} w \xi  +    \sigma^\ast\gamma_{42}\left( w+ \xi^{2 n +1}\right)  \\
\gamma_{23} \xi   - \sigma^\ast\gamma_{31} \left( w - \xi^{2n+1} \right)  & \gamma_{24} w \xi  -    \sigma^\ast\gamma_{41}\left( w - \xi^{2 n +1}\right)  \\
\end{array}
\right)\\
\end{split}   
\end{equation}
Here we remind that $\sigma^\ast \gamma_{ij}(\xi,w)=\gamma_{ij}(-\xi,w)$.

Let us give a name to the different zero modes, related to their charge under the $U(1)$ symmetry \eqref{U1generatorLaufer}:
$$\delta\phi_{13}\equiv \delta\phi_{B+}, \qquad
\delta\phi_{23}\equiv \delta\phi_{B-}, \qquad
\delta\phi_{14}\equiv \delta\phi_{B'+}, \qquad 
\delta\phi_{24}\equiv \delta\phi_{B'-}. $$
The modes  $\delta\phi_{B\pm}$ localize on the ideal
\begin{equation}
(\xi\,,\,\, w \pm \xi^{2n+1}) = (\xi\,,\,\, w) \:,
\end{equation}
while the modes $\delta\phi_{B'\pm}$ localize on the ideal
\begin{equation}
(\xi w\,,\,\, w \pm \xi^{2n+1}) = (\xi^{2n+2}\,,\,\, w\pm \xi^{2n+1})\:.
\end{equation}
Hence
\begin{eqnarray}
  \delta\phi_{B\pm} &\in& \mathbb{C}[\xi,w]/(\xi,w) \cong  \mathbb{C}  \\
  \delta\phi_{B'\pm} &\in& \mathbb{C}[\xi,w]/(\xi^{2n+2},w\pm \xi^{2n+1}) \cong \mathbb{C}[\xi]/(\xi^{2n+2}) \cong  \mathbb{C}^{2n+2}  
\end{eqnarray}
We then conclude that there are $1+ (2n+2)=2n+3$ hypermultiplets related to the off-diagonal blocks of $\delta\phi$. They have charge $1$ under the $U(1)$ symmetry generated by~\eqref{U1generatorLaufer}. This is understood form the brane point of view from the fact that these modes live on the intersection of the orientifold invariant branes with the $U(1)$ brane: they are bifundamentals of the intersecting branes, i.e. they have charge 1 under the $U(1)$ of the second brane (the first one is invariant and its group is projected out).

Summarizing, we obtained 
\begin{itemize}
\item \textbf{$\boldsymbol{n}$ modes} (and $n$ anti-modes) with charge ${\bf 2}$. 
\item \textbf{$\boldsymbol{2n+3}$ modes} (and $2n+3$ anti-modes) with charge ${\bf 1}$.  
\end{itemize}

This matches the known result that Laufer threefolds have the following GV invariants:
\be
n^{g=0}_{d=2} = n\,, \qquad\qquad n^{g=0}_{d=1} =2 n + 3 \,,
\ee
and zero for all other classes and genera.

\subsection{Morrison-Pinkham example}

A deformation of Laufer's threefold with $n=1$ was first discussed by \cite{pinkham}. Its defining equation is
\begin{equation}
 x^2 - z y^2 + (w +\lambda z) \left ( w^2 - z^{3} \right)  = 0 \:.
\end{equation}
with $\lambda$ a complex parameter.
This threefold has a singularity with a flop of length 2 at the origin ($x=y=z=w=0$). There is moreover a conifold singularity at the point $x=y=z-\lambda^2=w+\lambda^3=0$.

The corresponding D6-brane locus and tachyon matrix are given by
\begin{equation}\label{D6brlocusMPink}
(w+\lambda \xi^2 ) \, \xi ^2 \left(\xi ^{3}+w\right)\left(\xi ^{3}-w\right) = 0
\end{equation}
and
\begin{equation}\label{TachMPink}
T =\left(
\begin{array}{cccc}
 0 & \xi ^{3}+w & 0 & 0 \\
 \xi ^{3}-w & 0 & 0 & 0 \\
 0 & 0 & \xi  & 0 \\
 0 & 0 & 0 & (w+\lambda\xi^2) \xi  \\
\end{array}
\right) \:.
\end{equation}
One invariant brane has beed deformed with respect to Laufer.
The gauge symmetry on the D6-branes is like in Laufer, i.e. the $U(1)$ symmetry generated by \eqref{U1generatorLaufer} and living on the brane/image-brane system. 

The computation of the zero modes proceeds analogously to Laufer. The only significant difference occurs in the block $B$ of $\delta_gT$:
The modes  $\delta\phi_{B'\pm}$ now localize on the ideal
\begin{equation}
(\xi w + \lambda \xi^3 \,,\,\, w \pm \xi^{3}) = (\xi^3(\lambda\mp \xi)\,,\,\, w\pm \xi^{3})
\end{equation}
Hence, around $\xi=w=0$ we have
\begin{eqnarray}
  \delta\phi_{B'\pm} &\in& \mathbb{C}[\xi,w]/(\xi^{3},w) \cong \mathbb{C}[\xi]/(\xi^{3}) \cong  \mathbb{C}^{3}  \:,
\end{eqnarray}
while around $\xi=\pm\lambda$ and $w=-\lambda^3$
\begin{eqnarray}
  \delta\phi_{B'\pm} &\in& \mathbb{C}[\xi,w]/(\xi\mp \lambda,w - \lambda^{3}) \cong   \mathbb{C}  \:.
\end{eqnarray}
All these modes have charge one with respect to the $U(1)$ group living on the brane/image-brane stack. There is also one mode at the origin with charge equal to two: this can be derived with the same computation done in Laufer's example, with $n=1$.

The deformation from Laufer's example (with $n=1$) to Morrison-Pinkham's example has separated one conifold point from the singularity at the origin\footnote{One can further deform the threefold going to a manifold with $1+5$ conifold points.}. 
If we work in the completion of the local ring of the singularity, meaning, we zoom in on the origin, then we essentially move this conifold point infinitely far away. In that case, the GV invariants are 
\be
n^{g=0}_{d=2} = 1\,, \qquad\qquad n^{g=0}_{d=1} =4 \,,
\ee
and zero for all other invariants.

\subsection{Brown-Wemyss's example}

In \cite{BrownWemyss}, a different example of flop of length two was discussed\footnote{If we set $\zeta=1$ and take $z\mapsto -z-w$ one obtains the form in \cite{BrownWemyss}.}: 
\be
x^2 - z y^2 - w(w^2-z^3-2\zeta wz^2-\zeta^2 w^2z) = 0
\ee
Again we can easily read that the orientifold plane sits at $z=0$ and that the D6-brane locus in the double cover $\xi^2=z$ is given by
\begin{equation}\label{D6brlocusBW}
w  \, \xi ^2 \left(w+\xi(\zeta w+\xi ^{2})\right)\left(w-\xi(\zeta w+\xi ^{2})\right) = 0 \:.
\end{equation}
The corresponding tachyon matrix is
\begin{equation}\label{TachBW}
T =\left(
\begin{array}{cccc}
 0 & \xi(\xi^{2}+\zeta w)+ w & 0 & 0 \\
 \xi(\xi^{2}+\zeta w)-w & 0 & 0 & 0 \\
 0 & 0 & \xi  & 0 \\
 0 & 0 & 0 & w\xi  \\
\end{array}
\right)\:.
\end{equation}

Again, the $U(1)$ gauge group on the D6-branes is given by \eqref{U1generatorLaufer}. The zero mode computation follows the steps seen in Section~\ref{Sec:Laufer}:
We have one zero mode $\delta\phi_{A\pm}$ of charge two localized on the ideal
\begin{equation}
(w\,,\,\, \xi^{2} + \zeta  w) = (w\,,\,\, \xi ) \:.
\end{equation}
Moreover we have one zero mode of charge one localized  on
\begin{equation}
(\xi\,,\,\, w \pm \xi(\xi^{2}+\zeta w)) = (\xi\,,\,\, w)
\end{equation}
and four zero modes of charge one  localized on the ideal
\begin{equation}
(\xi w\,,\,\, w \pm (\xi^{3} +\zeta \xi w  )) = (\xi^{4}\,,\,\, w\pm \xi^{3})\:.
\end{equation}
We see that we have found the same spectrum that we obtained in Laufer's example with $n=1$.
As noticed in \cite{BrownWemyss}, the GV invariants of this manifold coincide with those of Laufer's example. This was actually the point of  \cite{BrownWemyss}, i.e. showing an example of two different varieties with the same GV invariants; they conclude that GV do not determine flops.

\section{Non-resolvable singularities: An intriguing family of threefolds}

\subsection{A simple example}
Let us look at a simple example for concreteness. Take the following hypersurface:
\be
u v = z^2 - w^3\,.
\ee
For this model, we can write the following tachyon matrix:
\be
T = \bm z& w^2\\w&z \eem\,.
\ee
Now, we compute the zero modes up to homotopies:
\be
\varphi = \bm \varphi_1 & \varphi_2 \\ \varphi_3 & \varphi_4 \eem \sim \varphi + T\cdot g_{L} + g_{R} \cdot T\,.
\ee
We will organize the calculation as follows: First, we notice the obvious fact that the tachyon matrix can be split into a part proportional to the identity plus a matrix $M$:
\be
T = z\cdot \mathbb{1}_2+M  \qquad {\rm with} \quad M = \bm 0& w^2\\w&0 \eem\,.
\ee
Now, we will show that we can eliminate any $z$-dependence of $\varphi$. Suppose
\be
\varphi = z\, \varphi_a (w, z)+\varphi_b (w)\,,
\ee
where $\varphi_a$ is a function of both $w$ and $z$, and $\varphi_b$ is only a function of $w$. Then, by choosing the left and right transformations such that $g_L+g_R = -\varphi_a$, the transformation subtracts the $z \varphi_a$ part, albeit generating further $w$-dependent terms. The main point is we gauge-fix to a function of $w$:
\be
\varphi = \tilde \varphi_b(w)
\ee
where $\tilde \varphi_b$ is not necessarily equal to $\varphi_b$. Now, we can perform further transformations that respect this gauge. Those are necessarily of the form: $g:= g_L = -g_R$. In this way, the whole spectral analysis boils down to studying $z$-independent fluctuations of $M$ modulo the commutator:
\be
\delta M \sim \delta M+[g, M]\,, \quad g \in sl_2\,.
\ee
Here, we can impose the traceless condition on $g$, since the commutator will kill that trace part.
Note, that this can be regarded as the study of fluctuations of an adjoint Higgs field on a brane at $z=0$, as studied in \cite{Cecotti:2010bp}, and further compared to the tachyon picture in \cite{Collinucci:2014qfa}.
Now we simply compute the most general such commutator:
\be
\left[\bm g_0&g_+\\g_-&-g_0 \eem, M \right]= \bm w (g_+ w-g_-) & 2 g_0 w^2 \\ -2 g_0 w& w (w g_+-g_-) \eem\,.
\ee
For the $(1,1)$ and $(2,1)$ entries, we can eliminate any $w$-dependence by choosing $g_-$ and $g_0$. The other two entries are then spread over the $z=0$ locus, and are seven-dimensional. Therefore, we find two localized zero-modes. Hence, we predict that there is one hyper.
\be
n^{g=0}_{d=1} = 1\,.
\ee
Note, that we are in this way $\emph{defining}$ Gopakumar-Vafa invariants for singular varieties \emph{without any reference} to a small resolution, since the variety is not small-resolvable in K\"ahler way.
In this case, we know that our counting is correct, because we can compare it to the Higgs branch results of \cite{Closset:2020scj}, where it is established that there is one hypermultiplet for M-theory on this variety by using a convoluted duality with IIB on the same threefold, thereby obtaining an Argyres-Douglas theory of type $(A_1, A_2)$.

\subsection{T-brane data}

In the previous sections, we gave families of geometries for which the choice of tachyon matrix was very canonical. Starting from a diagonal tachyon, one can in principle modify it, by switching on a so-called \emph{T-brane} background, such that the geometry of the brane and its M-theory uplift do not change, but the spectrum is altered. For instance, if we take the conifold case, we could alter the tachyon as follows:
\be
T = \bm z-w & 0 \\ 0 & z+w \eem \longrightarrow T = \bm z-w & 1 \\ 0 & z+w \eem\,.
\ee
The characteristic polynomial does not change, however the $u(1)$ flavor algebra is broken. Even though we have what appears to be a pair of intersecting D6-branes, in reality we have a single D6-brane (with a single center of mass) with a reducible shape. The M-theory uplift will still be the conifold, however, the open string vev in the off-diagonal entry translates to a {\bf coherent state of M2-branes} wrapping the vanishing $\pp^1$.

In IIA, we can immediately see that there is no spectrum left. We first diagonalize the tachyon via a bifundamental transformation, which amounts to a series of Gaussian moves:
\be
T = \bm z-w & 1 \\ 0 & z+w \eem \sim \bm z^2-w^2 &0\\0&1 \eem \cong \bm z^2-w^2\eem \,.
\ee
The last equality means that complex can now be split into a direct sum of a single brane on $z^2-w^2$, and a trivial fully eliminated brane. With this one-by-one tachyon matrix, we clearly see that there is no spectrum that localizes in 5d.

The moral here is that, for all cases we have treated so far, we have always made a canonical choice for the tachyon such that the spectrum is as large as can be. From that choice, further T-brane backgrounds can be switched on in order to reduce the spectrum in a hierarchical way.

In this section, however, we will show a class of examples where there is no canonical choice, and hence the spectrum is not uniquely defined. These cases correspond to geometries that do not admit K\"ahler crepant resolutions. They are defined as hypersurfaces as follows:
\be
u v = z^p-w^q\,, \quad {\rm for}\quad (p, q) \quad {\rm coprime}\,.
\ee
Because the powers are coprime, the rhs cannot be factorized. This prevents us from setting up a small resolution of the form:
\be
\bm u & \ast \\ \ast & v \eem \cdot \bm s\\t \eem=0\,.
\ee
So there is a priori no vanishing 2-cycle on which to wrap a membrane. Or, at the very least, this 2-cycle is forced to collapse. In the 5d effective theory, this means that there might be a spectrum of hypers for which no mass can be switched on. In M-theory, we do not know how to compute the GV invariants for singular varieties directly, since all methods rely on a small resolution, or a \emph{noncommutative crepant resolution}. These examples admit neither.

Here, we have a family of possible tachyons:
\be
T = \bm z^a & w^b\\w^c&z^d \eem\,, \qquad {\rm with} \quad a+d=p\,,\quad b+c=q\,.
\ee
For each choice, the geometry is the same, but the spectrum will be different. This means that, for non-resolvable singularities, the geometry alone is not sufficient to fully specify the physical background. Extra \emph{T-brane} data is needed.
The calculation for the generic case is complicated, so we will focus on the following family:
\be
u v = z^{2} + w^{2 k+1}\:.
\ee
Type IIB string theory on this threefold is known to give rise to the Argyres-Douglas theory AD$[A_{1}, A_{2 k}]$. Now the non-trivial choices for the tachyon matrix are parametrized by an integer $q$ as follows:
\be
T = \bm z & w^q\\w^{2 k+1-q} & z \eem\,, \quad 1 \leq q \leq 2k\,.
\ee
Now we apply the same line of reasoning as in the previous section: We first acknowledge that we can eliminate all $z$-dependence from the fluctuation modes by an appropriate combination of left and right gauge transformations. Then, we parametrize all remaining gauge transformations that respect the $z$-dependent gauge:
\be
\left[\bm g_0&g_+\\g_-&-g_0 \eem, M \right]= \bm g_+ w^{2 k+1-q}- g_- w^q & 2 g_0 w^q \\ -2 g_0 w^{2 k+1-q}& -g_+ w^{2 k+1-q}+g_- w^q \eem\,,
\ee
whereby $M := T\big|_{z=0}$. Now we have two possible outcomes:
\be
n^{g=0}_{d=1} = \begin{cases}
      2k+1-q & \,\,\,\text{if}\ 2 q \geq2k+1 \\
      q & \,\,\,\text{if}\ 2 q \leq 2k+1
    \end{cases}
\ee
Hence, we see a hierarchy for the singular GV invariants that depends on the choice of T-brane data. The maximal GV invariant is achieved by the choice $q = k+1$, yielding $n^0_{1} = k$. This matches the Higgs branch result of \cite{Closset:2020scj}.

Because the variety must remain singular, such coherent states of vanishing M2-branes have a strong influence on the spectrum of the effective theory, thereby leading to this multiplicity of possible outcomes. This is an entirely new situation, because the geometry does not give the full picture, but requires extra `sheafy' data. This general phenomenon was explored in different contexts in \cite{Collinucci:2016hpz,Collinucci:2014taa,Collinucci:2014qfa}, where it was also found that low-energy spectra are influenced by M2-coherent states.

\section{Conclusions}
In this paper, we have introduced a new way to calculate genus zero Gopakumar-Vafa invariants for non-compact CY threefolds that are $\cc^*$-fibrations, or $\zz_2$-orbifolds of $\cc^*$-fibrations. Specifically, we have worked on threefolds that admit flops of length one and two.
Our method consists in exploiting the M/IIA-duality, and recasting the problem of counting BPS invariants from M2-branes wrapped on holomorphic curves, to counting open string modes between D6-branes with O6-planes.

We have focused on `simple flops', meaning cases that have only one vanishing curve. In the family of Reid's pagoda, we found that, even though there is only one such curve, there are many non-zero GV invariants. In the length-two flops, such as Laufer's examples, we have interpreted degree-two invariants as strings that acquire charge-two under a flavor group, due to the orientifolding.

We have also touched upon a non-simple flop, called the generalized conifold. However, it would be interesting to investigate this more systematically. The threefolds we have studied are one-parameter families of $A$- and $D$-type local K3's, whereby only one curve of the Dynkin diagram is kept shrunken. However, there are more general possibilities where one keeps an arbitrary subgraph of the Dynkin graph shrunken. It would be interesting to study this further.

Finally, we have touched upon threefolds that admit $\cc^*$-fibrations, but do not admit crepant K\"ahler resolutions. For such threefolds, one cannot easily speak of vanishing curves, unless one is willing to perform a non-K\"ahler resolution. 
So, from the point of view of the topological string on the threefold, it is difficult to define the curve counting problem. However, upon reducing to IIA with D6-branes, the problem of counting open string modes becomes entirely tractable. Especially, if we treat the branes as coherent sheaves. Our results clearly show that T-brane data are crucial in defining the GV invariants.
In this paper, we work on a class of examples as an invitation to define GV invariants for non-resolvable threefolds. This warrants further investigation.

\section*{Acknowledgments} 

We have benefited from fruitful discussions with Mario De Marco. A.C.~is a Research Associate of the Fonds de la Recherche Scientifique F.N.R.S.~(Belgium). The work of A.C.~is partially supported by IISN - Belgium (convention 4.4503.15), and supported by the Fonds de la Recherche Scientifique - F.N.R.S.~under Grant CDR J.0181.18. 
The work of R.V.~is partially supported by ``Fondo per la Ricerca di Ateneo - FRA 2018'' (UniTS). 
A.S. and R.V. acknowledge support by INFN Iniziativa Specifica ST\&FI.



\providecommand{\href}[2]{#2}

\end{document}